\documentclass{article}
\usepackage{spconf,amsmath,graphicx,hyperref}
\usepackage{amsmath}   
\usepackage{amssymb}   
\usepackage{amsfonts}  
\usepackage{booktabs} 
\usepackage{multirow}
\usepackage{graphicx}   
\usepackage{booktabs}   
\usepackage{makecell}  
\usepackage{mathtools} 
\usepackage{amsfonts}  
\usepackage{colortbl, xcolor}

\definecolor{custompurple}{RGB}{203,192,231}
\usepackage{graphicx}
\usepackage[caption=false,font=footnotesize]{subfig} 
\usepackage{xcolor}
\definecolor{fake_color}{RGB}{211,128,135}
\definecolor{real_color}{RGB}{101,139,188}

\usepackage{tikz}
\usepackage{pgfplots}
\pgfplotsset{compat=1.18}

\usepackage{threeparttable} 
\usepackage[dvipsnames,table]{xcolor}
\definecolor{fired}{RGB}{220,50,47}   
\definecolor{iceblue}{RGB}{28,134,238}

\title{WaveSP-Net: Learnable Wavelet-Domain Sparse Prompt Tuning\\ for Speech Deepfake Detection}

\name{Xi Xuan$^{1,*}$, Xuechen Liu$^{2}$, Wenxin Zhang$^{3,5}$, Yi-Cheng Lin$^{4}$, Xiaojian Lin$^{6}$, Tomi Kinnunen$^{1}$}
\address{$^{1}$ University of Eastern Finland 
         $^{2}$ National Institute of Informatics \\
         $^{3}$ University of Chinese Academy of Sciences  
         $^{4}$ National Taiwan University \\
         $^{5}$ University of Toronto 
         $^{6}$ Tsinghua University}

\begin{document}
\ninept

\maketitle
\begingroup
\renewcommand\thefootnote{*}
\footnotetext{Corresponding author (xi.xuan@uef.fi)}
\endgroup
\begin{abstract}

Modern front-end design for speech deepfake detection relies on full fine-tuning of large pre-trained models like XLSR. However, this approach is not parameter-efficient and may lead to suboptimal generalization to realistic, in-the-wild data types. To address these limitations, we introduce a new family of parameter-efficient front-ends that fuse prompt-tuning with classical signal processing transforms. These include FourierPT-XLSR, which uses the Fourier Transform, and two variants based on the Wavelet Transform: WSPT-XLSR and Partial-WSPT-XLSR. We further propose WaveSP-Net, a novel architecture combining a Partial-WSPT-XLSR front-end and a bidirectional Mamba-based back-end. This design injects multi-resolution features into the prompt embeddings, which enhances the localization of subtle synthetic artifacts without altering the frozen XLSR parameters. Experimental results demonstrate that WaveSP-Net outperforms several state-of-the-art models on two new and challenging benchmarks, Deepfake-Eval-2024 and SpoofCeleb, with low trainable parameters and notable performance gains. The code and models are available online \footnote{https://github.com/xxuan-acoustics/WaveSP-Net}.

\end{abstract}


\vspace{-7pt}
\section{Introduction}

Speech deepfake detection (SDD) is the task of identifying artificially generated or manipulated speech audio, distinguishing it from 
bonafide human speech. This capability is critical for protecting speaker verification systems from various attacks, including speech synthesis, voice conversion, and voice cloning. Remarkable progress in SDD has been made on both front-end features~\cite{zhang2025multi, guo2024audio,ref43} and back-end models ~\cite{xuan2025fakemamba,elkheir25_interspeech}, achieving promising detection results especially on intra-domain settings. However, generalization to diverse unseen domains remains a major challenge; real-world settings require generalization to \emph{new} domains that may include unseen attacks, speech codecs, and audio compression formats~\cite{muller22_interspeech,xuan2024efficient}.

As in other detection tasks, the choice of front-end features for SDD is critically important. Existing front-ends can be broadly categorized into digital signal processing (DSP) and self-supervised learning (SSL) based approaches, each offering distinct advantages for cross-domain generalization. On one hand, the former includes methods such as short-time Fourier transform,  linear-frequency cepstral coefficients, and constant-Q transform 
~\cite{sahidullah2015comparison,ref5,ref6} aimed at capturing time-frequency characteristics using fixed transforms. On the other hand, 
modern data-driven SSL front-ends leverage foundational speech models such as XLSR~\cite{ref19} and Wav2Vec 2.0~\cite{baevski2020wav2vec} to extract information-rich features. 
SSL front-ends are typically fine-tuned on the new domain~\cite{ref13,ref14,xuan2024conformer,ref15, xuan25_spsc, ZHANG2026132741, li2026fast}.

\begin{figure*}[!t]
  \centering
  \includegraphics[width=0.83\textwidth, keepaspectratio]{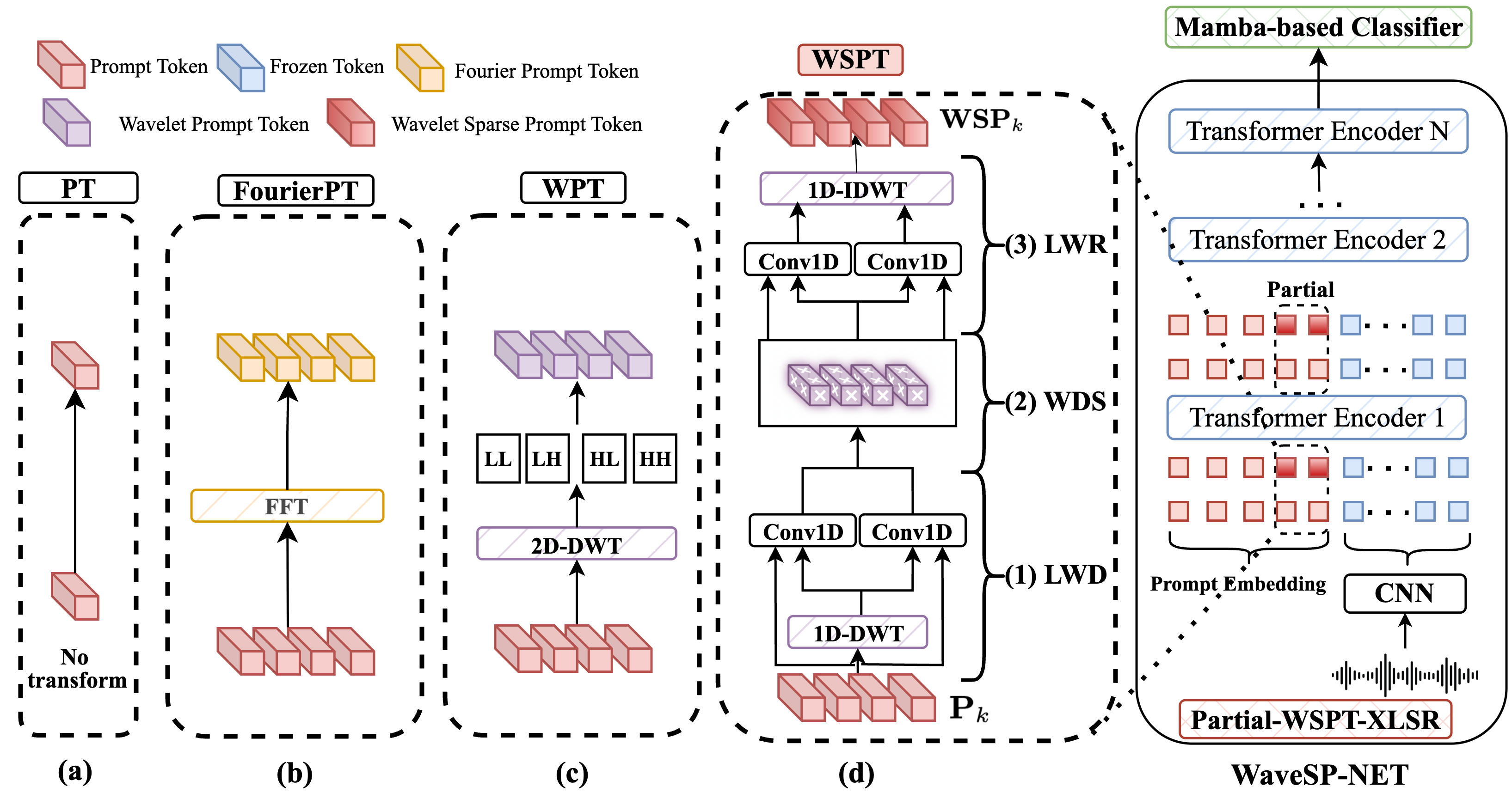}
   \caption{\footnotesize Overview of the WaveSP-Net architecture. The figure illustrates five different XLSR-based front-end variants: (a) PT-XLSR, (b) FourierPT-XLSR, (c) WPT-XLSR, (d) WSPT-XLSR and Partial-WSPT-XLSR. The proposed WaveSP-Net (rightmost panel) integrates a Partial-WSPT-XLSR front-end (bottom right) with a Mamba-based classifier (top right). (FFT: Fast Fourier Transform; DWT: Discrete Wavelet Transform; IDWT: Inverse Discrete Wavelet Transform)}
  \label{fig:wspt2}
  \vspace{-7pt}
\end{figure*}

While SSL front-ends typically achieve better detection performance over conventional models, they are computationally demanding and parameter-heavy, particularly with large SSL models with millions of parameters \cite{10832361}. As a data-driven technique, they are also prone to overfitting. To address these challenges, parameter-efficient fine-tuning (PEFT)~\cite{ding2023parameter} has emerged as a practical solution. PEFT refers to a broad family of methods aimed at adapting a foundation model to new domains while keeping the number of parameters requiring updating small. For instance, \cite{wu2024adapter} proposed intra-block and cross-block adapters to capture multi-level discriminative spoofing cues, 
whereas \cite{laakkonen2025generalizable} integrated LoRA adapters into the self-attention heads of XLSR-AASIST~\cite{jung2022aasist}, combined with meta-learning~\cite{vettoruzzo2024advances}. 

In this study, we focus on a particular promising PEFT approach, \textbf{prompt tuning} (PT)~\cite{lester-etal-2021-power}. As the name suggests, PT was originally introduced for modulating the behavior of large language models (LLMs) 
by providing them with additional ``instructions'' about a new task. This way, an \emph{existing} model can be reused for new tasks without the need for retraining. While this is the conceptual idea, the instructions---or \emph{prompt tokens}---are not actually hand-crafted text inputs, but additional model parameters that are optimized for the new task or domain.  
This makes PT widely applicable beyond LLMs as a generic PEFT method. Concretely, one freezes the original model, prepends the prompt parameters to selected parts of the model, and updates only them. The number of the prompt token parameters is typically a tiny fraction of the total parameter count, making PT a highly parameter-efficient solution. 

Despite its parameter efficiency and potential to improve domain generalization, PT has received surprisingly little attention in SDD. In~\cite {oiso24_interspeech}, the authors introduced a plug-in PT method for test-time domain adaptation to mitigate domain gaps with minimal target data and computational overhead. Our work (Fig.~\ref{fig:wspt2}) contributes to the recent line of research on advanced PT methods that enrich or constrain the structure of the prompt embeddings. In contrast to vanilla unstructured PT ~\cite{oiso24_interspeech} (leftmost block in Fig.~\ref{fig:wspt2}), prior work has used Fourier~\cite{zeng2024visual} and discrete wavelet~\cite{xie2025detect} transforms for this purpose. The key idea of our new approach (rightmost block in Fig.~\ref{fig:wspt2}) is to use the wavelet transform \cite{mallat2002theory} to enhance the prompt embeddings through a sparse transform-domain representation. As we demonstrate on the two recent and very challenging Deepfake-Eval-2024 (DE24)~\cite{chandra2025deepfakeeval2024} and large-scale SpoofCeleb~\cite{jung2025spoofceleb} benchmarks, our approach helps substantially in generalization. Our combined SDD solution, which combines XLSR front-end, the new wavelet prompting approach, and a recent Mamba-based back-end \cite{xuan2025fakemamba}, produces state-of-the-art results on both datasets.

\vspace{-7pt}
\section{Proposed Methods}
\label{sec:method}

This section details our proposed methods (see Fig.~\ref{fig:wspt2}). We first introduce three novel XLSR front-end variants that utilize classical DSP transforms to enrich prompt embedding representations. The existing and proposed PT methods are combined with a powerful Mamba-based \cite{xuan2025fakemamba} classifier. We dub our proposed architecture as \textbf{WaveSP-Net}. 

\vspace{-\baselineskip}

\subsection{FourierPT-XLSR}

Our first inspirations originate from a recently proposed PT approach known as visual Fourier prompt tuning (VFPT) \cite{zeng2024visual}. It adapts large transformers by augmenting fast Fourier transform (FFT) features into prompt embeddings, leading to strong results in vision tasks. We directly adopt this idea to SDD as a novel PEFT method, as illustrated in (Fig.~\ref{fig:wspt2}(b)). Our choice for the (frozen) SSL front-end is XLSR, 
given its competitive performance \cite{xin2022investigating}. We term the resulting FFT-based PT front-end as \textbf{FourierPT-XLSR}. 
\vspace{-5pt}
\subsection{WSPT-XLSR \& Partial-WSPT-XLSR}

Compared to the FFT based on non-localized sine and cosine bases and with uniform time-frequency tiling, the wavelet transform~\cite{arneodo1988wavelet} provides joint time–frequency localization with adaptive resolution, yielding robustness to signals with abrupt changes. To this end, we propose to augment discrete wavelet transform (DWT) coefficients into the prompt embeddings---sequences composed of prompt tokens, i.e., feature vectors produced by the XLSR front-end. We hypothesize this will help enhance artifact-sensitive frequency bands to enable fine-grained feature updates with minimal computational overhead. Inspired by \cite{zeng2024visual} and \cite{xie2025detect}, we \emph{selectively} apply wavelet-domain feature enhancement to only \emph{partial} prompt tokens within the prompt embeddings, termed \textbf{Partial-WSPT-XLSR} (Fig.~\ref{fig:wspt2}(d)). 

In the following, we detail the proposed front-end, which involves processing prompt embeddings through wavelet-domain enhancement. During training, the XLSR front-end remains frozen, with only updates applied to the PT and wavelet domain parameters. 
Concretely, for each of the Transformer layers $k \in \{1,\ldots,\ell\}$ in XLSR, we introduce $p$ additional learnable prompt tokens $\mathbf{P}_k \in \mathbb{R}^{p\times d}$,
where $d$ denotes the hidden dimension. Hence, each prompt token can be viewed as an additional 'input' 
with the same dimensionality as the features produced by XLSR. Note that each of the $\ell$ Transformer layers has its own set of parameters. During training, the prompt tokens are optimized; during inference, they are fixed 
and act as additional virtual inputs that 
guide the model. 

\textbf{The essence of our proposed method is to enforce additional structure to the prompt tokens through wavelet-domain processing.} To be specific, note that the vanilla prompt tokens described above are merely additional parameters described by unconstrained $d$-dimensional vectors optimized using any gradient-based method. Since the XLSR features themselves, however, are descriptors of a highly structured acoustic signal, we hypothesize that imposing additional structure to the prompt tokens themselves could lead to a more parameter-efficient model structure. In addition to their other benefits, wavelets are known for their ability to approximate prominent signal features (low-pass structure) and separate it from details (high-pass structure) such as noise. Concretely, our method transforms a pre-selected number of original tokens to a wavelet domain for additional processing, and combines these wavelet-domain processed token parameters with the original unprocessed ones.

\noindent\textbf{(1) Learnable Wavelet Decomposition (LWD).} 
Learnable wavelet transforms have used earlier in other applications such as compression of neural networks~\cite{wolter2020neural}, which is capable of dynamically adapting to different frequency domain signal characteristics. Inspired by this, we propose LWD. As illustrated in Fig.~\ref{fig:wspt2}, from each of the layer-specific prompt token sets $\mathbf{P}_k$, we select the last $m$ tokens $\mathbf{P}_k^{(p-m+1:p)}$ and transform them into \textit{wavelet sparse prompt} tokens $\mathbf{WSP}_k \in \mathbb{R}^{m\times d}$. 
In wavelet analysis~\cite{mallat2002theory}, a signal is separated into two complementary components: a low-frequency part that captures the overall, coarse structure of the input,  and a high-frequency part that captures the fine-grained detail information. This decomposition is performed using a pair of \textit{analysis filters}, denoted by $F_0$ (low-pass) and $F_1$ (high-pass). While in DSP applications, these filters are typically selected from a set of preset 'library' wavelets (e.g., Haar or Daubechies), in our model, they are \emph{learnable}~\cite{wolter2020neural}; the filter coefficients are optimized during PT training, allowing the model to adaptively extract coarse and fine information that is most useful for detecting deepfake speech.

\noindent\textbf{(2) Wavelet Domain Sparsification (WDS).}
After 1D discrete wavelet transform, the resulting low- and high-frequency coefficients are stacked into a single representation. However, the high and low frequency representations are located in a dense feature space, which compromises the computational efficiency and degrades the model's discriminative ability. To make learning more efficient and robust, we randomly select only a fraction of the feature positions to update, following the principle of sparse representations in compressed sensing~\cite{donoho2006compressed,bilican2025exploring}.  This stochastic sparsification in this architecture acts as an implicit regularizer: it reduces redundancy, helps prevent overfitting, and strengthens resistance to noise.

\noindent\textbf{(3) Learnable Wavelet Reconstruction (LWR).}
Finally, the processed wavelet-domain features are recombined into complete token representations using \textit{synthesis filters}, denoted $H_0$ (low-pass) and $H_1$ (high-pass), are designed to invert the earlier decomposition.  The analysis and synthesis filters are jointly learned, allowing the model to faithfully reconstruct the original prompt tokens while emphasizing their most prominent coefficients. The result is a set of compact, expressive, and robust, enhanced prompt tokens.

\vspace{-5pt}
\subsection{WaveSP-Net}
After computing $\mathbf{WSP}_k$, we obtain the final prompt representation that integrates both enhanced and untransformed tokens:
\begin{equation}
\tilde{\mathbf{P}}_k=[\mathbf{P}_k^{(1:p-m)},\mathbf{WSP}_k]\in\mathbb{R}^{p\times d},
\end{equation}
where $k \in \{1,\ldots,\ell\}$ indexes the transformer layer, $p$ denotes the total number of prompt tokens per layer, and $0 \leq m \leq p$. Thus, $\tilde{\mathbf{P}}_k$ has the same shape as the original prompt $\mathbf{P}_k$, but the improved wavelet-based representations replace its last $m$ positions. Next, the modified prompt tokens are inserted into the transformer computation. At layer $k$, the input is the concatenation of the processed prompt tokens $\tilde{\mathbf{P}}_k$ and the previous layer’s embeddings $\mathbf{E}_{k-1}$.  
Passing these through the $k$-th transformer layer $L_k(\cdot)$ yields:
\begin{equation}
[\mathbf{Z}_k, \mathbf{E}_k] = L_k([\tilde{\mathbf{P}}_k, \mathbf{E}_{k-1}]), \quad k = 1, 2, \ldots, \ell
\end{equation}
where $\mathbf{Z}_k$ represents the transformed prompt outputs at layer $k$, and  $\mathbf{E}_k$ is the updated sequence embedding output by the same layer. Finally, the output of the transformer final layer $I = [\mathbf{Z}_l, \mathbf{E}_l]$ will be sent to the Mamba-based classifier ~\cite{xuan2025fakemamba}. The Mamba architecture is well-suited for high-dimensional wavelet domain representations because it effectively captures long-range temporal dependencies while maintaining linear time complexity. During training, only the prompt embeddings, learnable wavelet filters, and Mamba-based classifier parameters are updated while keeping the XLSR backbone frozen, ensuring parameter efficiency.
\begin{table}[t]
    \centering
    \footnotesize
    \setlength{\tabcolsep}{4pt}
    \renewcommand{\arraystretch}{1.25}
    \caption{\footnotesize Deepfake-Eval-2024 and SpoofCeleb benchmark results for three proposed front-ends: FourierPT-XLSR, WSPT-XLSR, and Partial-WSPT-XLSR, each combined with a shared Mamba-based classifier. The best results are in \textbf{bold}. The 95\% parametric confidence intervals for EER are shown in parentheses.}
    \vspace{0.1cm}
    \label{tab:1}
    \resizebox{\columnwidth}{!}{
    \begin{tabular}{@{}l c c c c@{}}
        \Xhline{1.4pt}
        \textbf{Model} & \multicolumn{4}{c}{\textbf{Deepfake-Eval-2024}} \\
        \cmidrule(lr){2-5}
        & \textbf{EER (\%)} $\downarrow$ & \textbf{ACC (\%)} $\uparrow$ & \textbf{F1 (\%)} $\uparrow$ & \textbf{AUC (\%)} $\uparrow$ \\
        \Xhline{0.8pt}
        \textbf{FourierPT-XLSR}      & 16.58 ($\pm$ 0.52) & 83.42 & 79.53 & 90.35 \\
        \textbf{WSPT-XLSR}           & 13.15 ($\pm$ 0.47) & 86.85 & 83.84 & 93.33 \\
        \textbf{Partial-WSPT-XLSR}   & \textbf{10.58} ($\pm$ 0.43) & \textbf{89.42} & \textbf{86.35} & \textbf{94.26} \\
        \Xhline{0.8pt}
        \textbf{Model} & \multicolumn{4}{c}{\textbf{SpoofCeleb}} \\
        \cmidrule(lr){2-5}
        & \textbf{EER (\%)} $\downarrow$ & \textbf{ACC (\%)} $\uparrow$ & \textbf{F1 (\%)} $\uparrow$ & \textbf{AUC (\%)} $\uparrow$ \\
        \Xhline{0.8pt}
        \textbf{FourierPT-XLSR}      & 0.23 ($\pm$ 0.06) & 99.84 & 99.87 & 99.86 \\
        \textbf{WSPT-XLSR}           & 0.19 ($\pm$ 0.06) & 99.89 & 99.92 & 99.91 \\ 
        \textbf{Partial-WSPT-XLSR}   & \textbf{0.13} ($\pm$ 0.04) & \textbf{99.87} & \textbf{99.93} & \textbf{99.99} \\
        \Xhline{1.4pt}
    \end{tabular}}
    \vspace{-0.5cm}
\end{table}

\vspace{-6pt}
\section{Experimental Setup}
\subsection{Dataset \& Metrics}

Our experiments use two benchmarks: Deepfake-Eval-2024 (DE24) ~\cite{chandra2025deepfakeeval2024} and SpoofCeleb~\cite{jung2025spoofceleb}. To evaluate deepfake detector generalization, we train and evaluate on each dataset separately following its official protocol. For Deepfake-Eval-2024 \footnote{https://huggingface.co/datasets/nuriachandra/Deepfake-Eval-2024}, we follow~\cite{chandra2025deepfakeeval2024} and preprocess the audio subset by chunking long clips into 4-second segments. Spanning 88 web domains and 42 languages, DE24 includes audio samples with varying acoustic conditions, thereby subjecting the detector to strictly unseen attacks and complex distribution shifts. For SpoofCeleb, we follow the established protocol\footnote{https://www.jungjee.com/spoofceleb/}: the attacks included in the training are A01-A10, while the ones for evaluations are A15-A23. For more details, please refer to \cite{jung2025spoofceleb}. Performance is reported with EER, AUC, F1, and accuracy (ACC). We also report 95\% parametric confidence intervals for EER following~\cite{bengio04_odyssey}: \(\text{EER} \pm \sigma \cdot Z_{\alpha/2}\), where \(Z_{\alpha/2} = 1.96\), \(\sigma = 0.5 \sqrt{\text{EER}(1 - \text{EER}) \left({n_r + n_f})/({n_r n_f}\right)}\), where \(n_r\) and \(n_f\) denote the number of real and fake samples, respectively.
\vspace{-5pt}
\subsection{Implementation Details} Each of the experiments is conducted on a standalone Tesla V100 GPU with a fixed random seed. Audio samples are down-sampled to 16 kHz and padded or cropped to 4 seconds, before being processed by the XLSR-300M SSL feature extractor\footnote{https://huggingface.co/facebook/wav2vec2-xls-r-300m} to produce 2D features of size $(201, 1024)$.
For PT, FourierPT, and WSPT, we use $p=10$ prompt tokens; for WPT and Partial-WSPT, these tokens consist of four wavelet-based and six regular tokens. The sparsity ratio is $\rho=0.1$. Hyperparameter sensitivity to prompt token and sparsity ratio configurations is analyzed in Section~\ref{sec:ablation}. The Mamba-based classifier comprises 12 Mamba-based blocks. Training uses a dropout of 0.1, batch size of 16, learning rate of $5\times 10^{-4}$, and the Adam optimizer. Models are trained with cross-entropy loss for up to 100 epochs, with early stopping when development loss plateaus for seven consecutive iterations. Models are selected from the checkpoint that yields the lowest EER on the development set.

\vspace{-5pt}
\section{Results and analysis}

\subsection{Framework with Three Novel XLSR Variants Front-Ends}

\begin{table}[t]
\centering
\footnotesize
\setlength{\tabcolsep}{4pt}
\renewcommand{\arraystretch}{1.25}
\caption{\footnotesize Comparison with SOTA single systems on the Deepfake-Eval-2024 benchmark. The best results are in \textbf{bold}, and the second-best are \underline{underlined}. The 95\% parametric confidence intervals for EER are shown in parentheses. BCM denotes the Best Commercial Model.}
\label{tab:2}
\vspace{3pt}
\resizebox{\columnwidth}{!}{
\begin{tabular}{@{}l c c c c c@{}}
\Xhline{1.4pt}
\textbf{Model} & \textbf{Params (\% of Total)} & \textbf{EER (\%)} $\downarrow$ & \textbf{ACC (\%)} $\uparrow$ & \textbf{F1 (\%)} $\uparrow$ & \textbf{AUC (\%)} $\uparrow$ \\
\Xhline{0.8pt}
AASIST~\cite{chandra2025deepfakeeval2024} & 0.3M & 16.99 ($\pm$ 0.52) & 83.60 & 77.80 & 90.60 \\
RawNet2~\cite{chandra2025deepfakeeval2024} & 18M & 20.91 ($\pm$ 0.56) & 81.70 & 86.00 & 87.60 \\
P3~\cite{chandra2025deepfakeeval2024} & 317M & 15.38 ($\pm$ 0.50) & 85.50 & 81.00 & 92.00 \\
XLS\text{-}R\text{-}1B~\cite{ge2025post} & 965M & \underline{11.85 ($\pm$ 0.45)} & 86.83 & \textbf{89.43} & \textbf{94.35} \\
BCM~\cite{chandra2025deepfakeeval2024} & - & - & \underline{89.00} & \underline{87.00} & 93.00 \\
\midrule
PT\text{-}XLSR & 4.145M & 20.40 ($\pm$ 0.56) & 79.60 & 77.19 & 90.21 \\
WPT\text{-}XLSR & 4.145M & 14.39 ($\pm$ 0.49) & 85.61 & 81.01 & 91.29 \\
\Xhline{1.4pt}
\textbf{WaveSP-Net} & 4.146M \textbf{(1.298\%)} & \textbf{10.58} ($\pm$ 0.43) & \textbf{89.42} & 86.35 & \underline{94.26} \\
\Xhline{1.4pt}
\end{tabular}}
\vspace{-0.5cm}
\end{table}

\vspace{-6pt}
\begin{table}[t]
    \centering
    \footnotesize
    \setlength{\tabcolsep}{4pt}
    \renewcommand{\arraystretch}{1.25}  
    \caption{\footnotesize Comparison with SOTA single systems on the SpoofCeleb benchmark. The best results are in \textbf{bold}. The 95\% parametric confidence intervals for EER are shown in parentheses. Params denotes trainable parameters.}
    \label{tab:3}
    \vspace{3pt}
    \resizebox{\columnwidth}{!}{
    \begin{tabular}{@{}l c c c c c@{}}
        \Xhline{1.4pt}
        \textbf{Model} & \textbf{Params (\% of Total)} & \textbf{EER (\%)} $\downarrow$ & \textbf{ACC (\%)} $\uparrow$ & \textbf{F1 (\%)} $\uparrow$ & \textbf{AUC (\%)} $\uparrow$ \\
        \Xhline{0.8pt}
        AASIST~\cite{jung2025spoofceleb}         & 0.3M   & 2.37 ($\pm$ 0.16) & 71.38                      & 81.25                      & 83.56             \\
        RawNet2~\cite{jung2025spoofceleb}        & 18M    & 1.12 ($\pm$ 0.11) & 87.23                      & 88.92                      & 92.14           \\
        \midrule
        PT\text{-}XLSR  & 4.145M & 0.26 ($\pm$ 0.06) & 99.74  & 99.85 & 99.93             \\
        WPT\text{-}XLSR & 4.145M & 0.15 ($\pm$ 0.04) & 99.85  & 99.92   & 99.97 \\
        \Xhline{1.4pt}
        \textbf{WaveSP-Net}   & 4.146M \textbf{(1.298\%)} & \textbf{0.13} ($\pm$ 0.04) & \textbf{99.87} & \textbf{99.93} & \textbf{99.99} \\
        \Xhline{1.4pt}
    \end{tabular}}
    \vspace{-0.5cm}
\end{table}

\begin{figure}[t]
    \centering
    \setlength{\fboxrule}{0.8pt}
    \setlength{\fboxsep}{0pt}
    \footnotesize  

    \subfloat[FourierPT-XLSR\label{fig:tsne-wpt}]{
        \fbox{\includegraphics[width=0.31\linewidth]{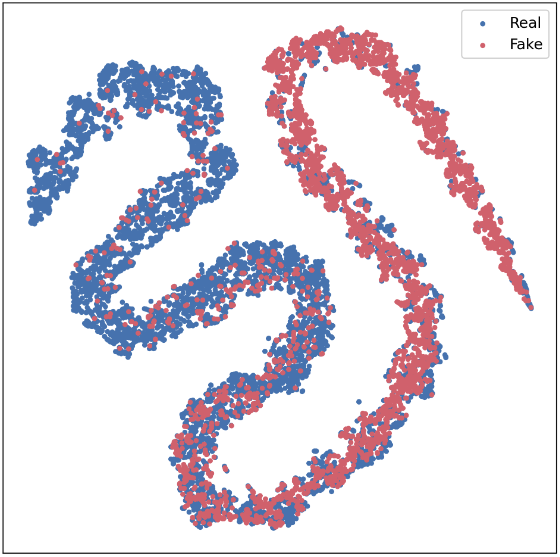}}
    }\hfill
    \subfloat[WSPT-XLSR\label{fig:tsne-fourierpt}]{
        \fbox{\includegraphics[width=0.31\linewidth]{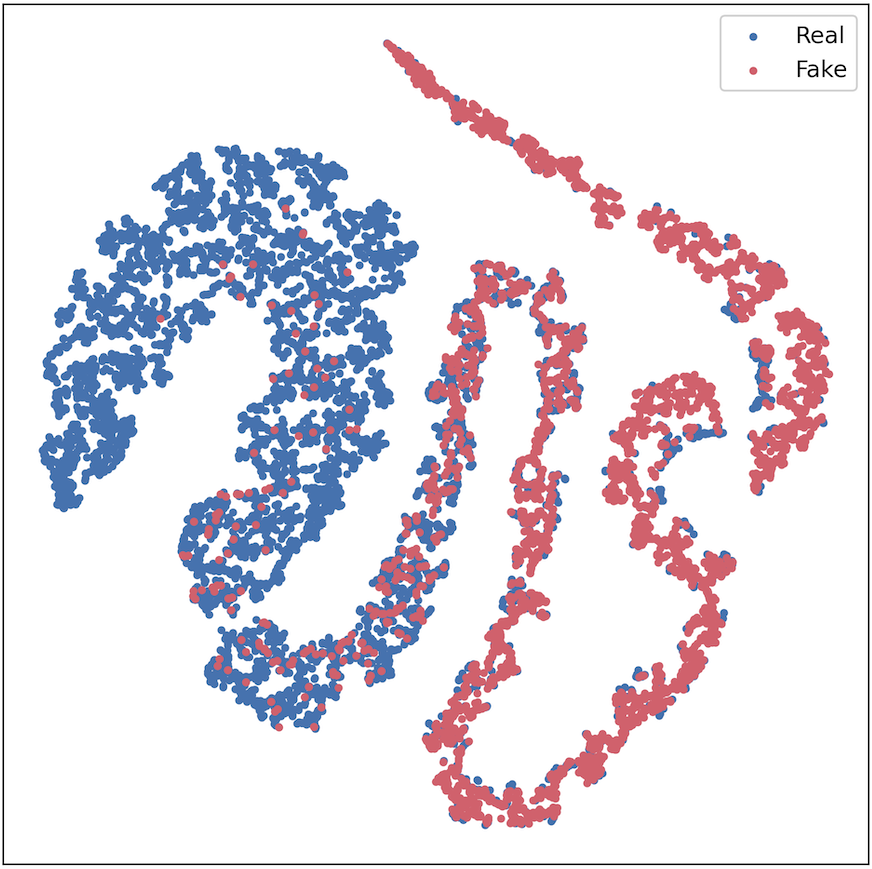}}
    }\hfill
    \subfloat[Partial-WSPT-XLSR\label{fig:tsne-ppwt}]{
        \fbox{\includegraphics[width=0.31\linewidth]{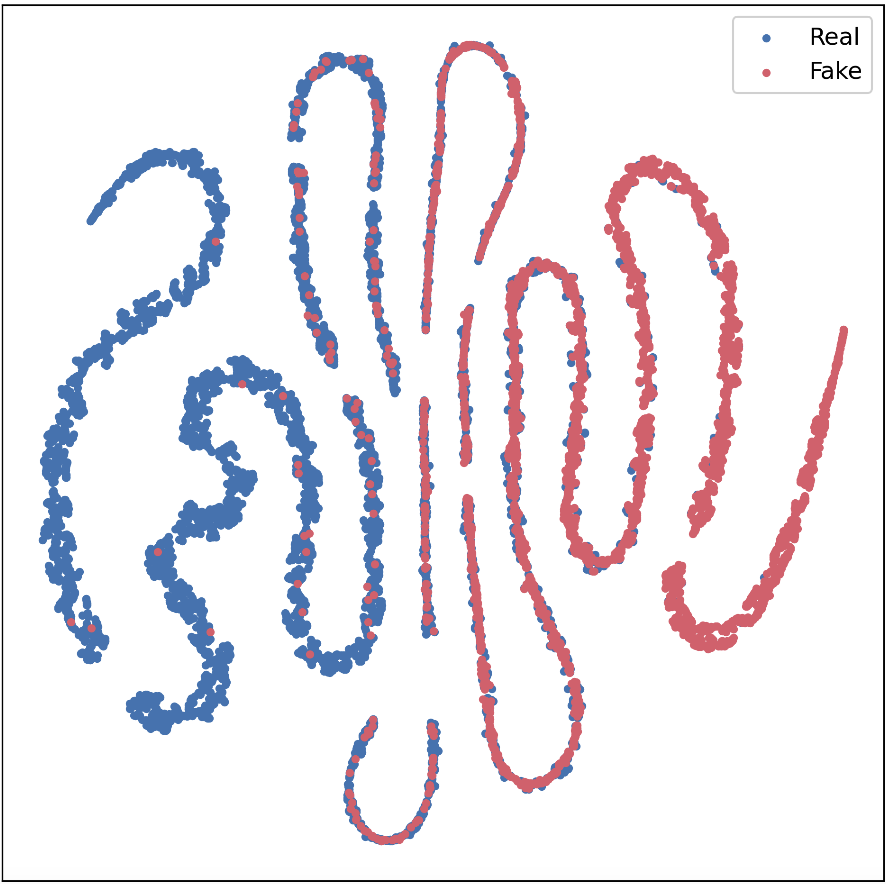}}
    }

    \vspace{1mm}  
    
    \begin{minipage}{\linewidth}
        \centering
        \textcolor{real_color}{\rule{1.0em}{1.0ex}}~Real\quad
        \textcolor{fake_color}{\rule{1.0em}{1.0ex}}~Fake
    \end{minipage}

    \vspace{-5pt}

    \caption{\footnotesize 2D t-SNE visualization of the Deepfake-Eval-2024 test set.}
    \label{fig:tsne}
    \vspace{-0.5cm}
\end{figure}

Table~\ref{tab:1} shows the performance of our three proposed front-ends on the DE24 and SpoofCeleb benchmarks. The results clearly indicate that the wavelet-based front-ends, WSPT-XLSR and Partial-WSPT-XLSR, outperform the FourierPT-XLSR front-end. Specifically, Partial-WSPT-XLSR achieves the best results on both datasets, with the lowest EER of 10.58\% on DE24 and 0.13\% on SpoofCeleb, and the highest scores across accuracy, F1, and AUC metrics. This trend suggests that wavelet-based feature extraction is more effective at capturing discriminative characteristics than Fourier-based methods, likely due to its joint time-frequency analysis capabilities.

\vspace{-8pt}
\subsection{Comparison with SOTA Models on the Two Benchmark}

Table \ref{tab:2} compares WaveSP-Net against several SOTA single systems on the DE24 benchmark. The model achieves an EER of 10.58\%, representing a 10.72\% relative improvement over the leading XLS-R-1B and a 2.59\% accuracy gain, while requiring significantly fewer trainable parameters (only 1.298\% of total parameters). 

Table \ref{tab:3} presents performance comparisons on the SpoofCeleb benchmark, where WaveSP-Net achieves the lowest EER among compared models. The model attains an EER of 0.13\% (13.33\% relative improvement over WPT-XLSR), with ACC, F1, and AUC of 99.87\%, 99.93\%, and 99.99\%, respectively. The consistent performance across both datasets indicates the model's effectiveness in detecting synthetic speech artifacts.

\vspace{-8pt}
\subsection{Ablation \& Parameter Sensitivity Experiments}
\label{sec:ablation}

Table \ref{tab:ablation_results} provides a detailed ablation study on the core components of the WaveSP-Net. Our results indicate that removing any core component leads to a notable performance degradation, with WDS causing the most significant drop by relatively 35.54\% in EER. This highlights the critical role of the sparsity mechanism in filtering out noise. Additionally, replacing learnable wavelet filters with fixed ones also decreased performance, with a relative increase of 56.44\% in EER, validating that learnable wavelet filters are effectively co-optimized with the back-end to learn discriminative features. We also perform hyperparameter sensitivity analysis on two key parameters: sparsity ratio and the number of wavelet sparse prompt tokens, as shown in Table~\ref{tab:ablation_results}. 
Experimental results indicate that the optimal WaveSP-Net configuration consists of learnable wavelet filters, a sparsity ratio of 0.1, and four wavelet sparse prompt tokens.

\begin{table}[t]
    \centering
    \caption{\footnotesize Ablation \& Parameter Sensitivity Results for WaveSP-Net (Partial-WSPT-XLSR as front-end) on DE24 datasets. Best results are in bold. (WaveSP-Net settings: Learnable Wavelet Filters, 10\% Sparsity Ratio, and 4 Wavelet Sparse Prompt Tokens.)}
    \label{tab:ablation_results}
    \footnotesize  
    \vspace{5pt}
    \setlength{\tabcolsep}{4pt}
    \renewcommand{\arraystretch}{1.05}
    \scalebox{0.9}{
    \vspace{3pt}\begin{tabular}{@{}l c c c c@{}}
        \Xhline{1.4pt}
        & \textbf{EER (\%)} $\downarrow$ & \textbf{ACC (\%)} $\uparrow$ & \textbf{F1 (\%)} $\uparrow$ & \textbf{AUC (\%)} $\uparrow$ \\
        \Xhline{0.5pt}
\rule{0pt}{12pt}\textbf{WaveSP-Net} & \textbf{10.58} ($\pm$ 0.43) & \textbf{89.42} & \textbf{86.35} & \textbf{94.26}\rule[-4pt]{0pt}{0pt} \\[2pt]
\Xhline{0.8pt}
        \multicolumn{5}{c}{\textbf{Ablation1: Partial-WSPT-XLSR}} \\
        \Xhline{0.3pt}
        w/o LWD & 12.97 ($\pm$ 0.47) & 87.03 & 84.37 & 94.00 \\
        w/o WDS & 14.34 ($\pm$ 0.49) & 85.66 & 83.09 & 93.73 \\
        w/o LWR & 11.33 ($\pm$ 0.44) & 88.67 & 85.33 & 94.09 \\
        \Xhline{0.3pt}
        \multicolumn{5}{c}{\textbf{Ablation2: Fixed vs Learnable Wavelet Filters}} \\
        \Xhline{0.3pt}
        Fixed Filters & 16.55 ($\pm$ 0.51) & 83.45 & 79.63 & 90.36 \\
        \Xhline{0.8pt}
        \multicolumn{5}{c}{\textbf{Hyperparameter1: Sparsity Ratio}} \\
        \Xhline{0.3pt}
        0.5 & 12.42 ($\pm$ 0.46) & 87.58 & 84.49 & 93.44 \\
        0.7 & 13.84 ($\pm$ 0.48) & 86.16 & 83.31 & 93.56 \\
        0.9 & 12.73 ($\pm$ 0.46) & 87.27 & 84.28 & 93.75 \\
        \Xhline{0.8pt}
        \multicolumn{5}{c}{\textbf{Hyperparameter2: Wavelet Sparse Prompt Token}} \\
        \Xhline{0.3pt}
        2 & 11.23 ($\pm$ 0.44) & 88.77 & 84.31 & 93.82 \\
        6 & 14.86 ($\pm$ 0.49) & 85.14 & 81.04 & 91.03 \\
        8 & 12.65 ($\pm$ 0.46) & 87.35 & 84.50 & 93.88 \\
        10 & 13.15 ($\pm$ 0.47) & 86.85 & 83.84 & 93.33 \\
        \Xhline{1.4pt}
    \end{tabular}}
    \vspace{-0.5cm}
\end{table}
\vspace{-8pt}
\subsection{Visualization}

Fig.~\ref{fig:tsne} presents a 2D t-SNE visualization of the DE24 test set. In Figs.~\ref{fig:tsne}(a) and (b), FourierPT-XLSR and WSPT-XLSR show significant overlap between real (blue) and fake (red) samples, echoing their detection performance. In contrast, Partial-WSPT-XLSR, shown in Fig.~\ref {fig:tsne}(c), displays distinct, tight, and highly isolated clusters with minimal overlap. This visualization demonstrates that WaveSP-Net effectively learns highly discriminative features by focusing on sparse, informative features in the wavelet domain.

\vspace{-5pt}
\section{Conclusion}
This paper introduces WaveSP-Net, a novel speech deepfake detector that combines a Partial-WSPT-XLSR front-end with a bidirectional Mamba back-end. The core innovation lies in using learnable wavelet filters to enhance a sparse subset of prompt tokens adaptively. This approach turns out to be a parameter-efficient and effective solution for SDD. Our experiments indicate that WaveSP-Net outperforms other SOTA single systems on two new and challenging benchmarks, Deepfake-Eval-2024 and SpoofCeleb, achieving SOTA performance with low trainable parameters. This successful integration of classical signal processing transforms into our architecture prompts us to reconsider their role. We believe these findings will inspire new approaches that combine hand-crafted acoustic features with the power of large language models.

\vspace{-5pt}
\section{Acknowledgment}

This work was supported by the Finnish Doctoral Program Network in Artificial Intelligence (AI-DOC), project “Explainable Speech Deepfake Characterization” (Decision No. VN/3137/2024-OKM-6), and partially by the Research Council of Finland, project “SPEECHFAKES” (Decision No. 349605).

\vspace{-4pt}
\small
\bibliographystyle{IEEEtran}
\bibliography{refs}

@inproceedings{muller22_interspeech,
  title     = {{Does Audio Deepfake Detection Generalize?}},
  author    = {{Nicolas Müller and others}},
  year      = {{2022}},
  booktitle = {{Interspeech 2022}},
  pages     = {{2783--2787}},
  doi       = {{10.21437/Interspeech.2022-108}},
  issn      = {{2958-1796}},
}

@inproceedings{ref13,
  title     = {Channel-Wise Gated Res2Net: Towards Robust Detection of Synthetic Speech Attacks},
  author    = {Xu Li and others},
  year      = {2021},
  booktitle = {Interspeech 2021},
}

@inproceedings{xuan25_spsc,
  title     = {{Multilingual Source Tracing of Speech Deepfakes: A First Benchmark}},
  author    = {Xi Xuan and others},
  year      = {2025},
  booktitle = {{5th Symposium on Security and Privacy in Speech Communication}},
  pages     = {27--34},
  doi       = {10.21437/SPSC.2025-5},
}

@article{xuan2024conformer,
  title={Conformer-Based Speaker Recognition Model for Real-Time Multi-Scenarios},
  author={Xuan, Xi and others},
  journal={Computer Engineering and Applications},
  volume={60},
  number={7},
  pages={147--156},
  year={2024}
}

@inproceedings{laakkonen2025generalizable,
  title={Generalizable speech deepfake detection via meta-learned LoRA},
  author={Laakkonen, Janne and others},
  booktitle={ICASSP 2026-2026 IEEE International Conference on Acoustics, Speech and Signal Processing}
}

@inproceedings{wu2024adapter,
  title={Adapter learning from pre-trained model for robust spoof speech detection},
  author={Wu, Haochen and others},
  booktitle={Interspeech},
  year={2024}
}

@inproceedings{li2026fast,
  title={{FAST\_QR}: Fast, Accurate and Stable Quantile Regression for Time-Series Analysis via Adaptive {Huber} Smoothing},
  author={Li, Zhiyu and others},
  booktitle={2026 IEEE International Conference on Acoustics, Speech and Signal Processing (ICASSP)},
  year={2026},
  organization={IEEE}
}

@article{ZHANG2026132741,
title = {Robust rumor detection against noise},
journal = {Neurocomputing},
pages = {132741},
year = {2026},
issn = {0925-2312},
doi = {https://doi.org/10.1016/j.neucom.2026.132741},
author = {Wenxin Zhang and others}
}

@inproceedings{ref14,
  title     = {{Does Audio Deepfake Detection Generalize?}},
  author    = {{Nicolas Müller and others}},
  year      = {{2022}},
  booktitle = {{Interspeech 2022}},
  pages     = {{2783--2787}},
  doi       = {{10.21437/Interspeech.2022-108}},
  issn      = {{2958-1796}},
}

@article{mallat2002theory,
  title={A theory for multiresolution signal decomposition: the wavelet representation},
  author={Mallat, Stephane G},
  journal={IEEE transactions on pattern analysis and machine intelligence},
  pages={674--693},
  year={2002},
  publisher={Ieee}
}

@article{ref15,
  title={The role of long-term dependency in synthetic speech detection},
  author={Li, C and others},
  journal={IEEE Signal Processing Letters},
  year={2022}
}

@inproceedings{ref43,
  author={Zhang, Q. and others},
  title={{Audio deepfake detection with self-supervised XLS-R and SLS classifier}},
  booktitle={Proceedings of the 32nd ACM International Conference on Multimedia},
  year={2024}
}

@inproceedings{elkheir25_interspeech,
  title     = {{BiCrossMamba-ST: Speech Deepfake Detection with Bidirectional Mamba Spectro-Temporal Cross-Attention}},
  author    = {{Yassine {El Kheir} and others}},
  year      = {{2025}},
  booktitle = {{Interspeech 2025}},
  doi       = {{10.21437/Interspeech.2025-527}},
  issn      = {{2958-1796}},
}

@inproceedings{xie2025detect,
  title={Detect All-Type Deepfake Audio: Wavelet Prompt Tuning for Enhanced Auditory Perception},
  author={Xie, Yuankun and others},
  booktitle = {Proceedings of the AAAI Conference on Artificial Intelligence},
  year={2026}
}

@article{donoho2006compressed,
  title={Compressed sensing},
  author={Donoho, David L},
  journal={IEEE Transactions on information theory},
  year={2006}
}

@article{vettoruzzo2024advances,
  title={Advances and challenges in meta-learning: A technical review},
  author={Vettoruzzo, Anna and others},
  journal={IEEE transactions on pattern analysis and machine intelligence},
  year={2024}
}

@article{ding2023parameter,
  title={Parameter-efficient fine-tuning of large-scale pre-trained language models},
  author={Ding, Ning and others},
  journal={Nature machine intelligence},
  year={2023}
}

@misc{chandra2025deepfakeeval2024,
      title={Deepfake-Eval-2024: A Multi-Modal In-the-Wild Benchmark of Deepfakes Circulated in 2024}, 
      author={Nuria Alina Chandra and others},
      year={2025},
      eprint={2503.02857},
      archivePrefix={arXiv},
      primaryClass={cs.CV},
      url={https://arxiv.org/abs/2503.02857}, 
}

@inproceedings{bengio04_odyssey,
  title     = {A statistical significance test for person authentication},
  author    = {Samy Bengio and Johnny Mariéthoz},
  year      = {2004},
  booktitle = {The Speaker and Language Recognition Workshop (Odyssey 2004)},
  pages     = {237--244},
}

@article{jung2025spoofceleb,
  title={SpoofCeleb: Speech deepfake detection and SASV in the wild},
  author={Jung, Jee-weon and others},
  journal={IEEE Open Journal of Signal Processing},
  year={2025}
}

@inproceedings{zhang2025multi,
  author    = {Zhang, Kai and others},
  title     = {{Multi-View Collaborative Learning Network for Speech Deepfake Detection}},
  booktitle = {Proceedings of the AAAI Conference on Artificial Intelligence},
  year      = {2025},
  pages     = {1075--1083}
}

@inproceedings{sahidullah2015comparison,
  title={A comparison of features for synthetic speech detection},
  author={Sahidullah, Md and others},
  booktitle={Proceedings of Interspeech},
  year={2015}
}

@inproceedings{wolter2020neural,
	title = {Neural {Network} {Compression} via {Learnable} {Wavelet} {Transforms}},
	author = {Wolter, Moritz and others},
	booktitle = {International {Conference} on {Artificial} {Neural} {Networks}},
	year = {2020},
	doi = {10.1007/978-3-030-61616-8_4},
	organization = {Springer},
}

@article{bilican2025exploring,
  title={Exploring Sparsity for Parameter Efficient Fine Tuning Using Wavelets},
  author={Bilican, Ahmet and others},
  journal={arXiv preprint arXiv:2505.12532},
  year={2025}
}

@article{arneodo1988wavelet,
  title={Wavelet transform of multifractals},
  author={Arneodo, A and others},
  journal={Physical review letters},
  volume={61},
  number={20},
  pages={2281},
  year={1988},
  publisher={APS}
}

@inproceedings{xuan2024efficient,
  title={Efficient Real-Time Multi-Scenario Speaker Recognition with Mel-Spectrogram-Based Hybrid TDNN for Edge System},
  author={Xuan, Xi and others},
  booktitle={Interspeech 2024 - Young Female Researchers in Speech Workshop}
}

@inproceedings{zeng2024visual,
  title={Visual fourier prompt tuning},
  author={Zeng, Runjia and others},
  booktitle = {Advances in Neural Information Processing Systems (NeurIPS)},
  year={2024}
}

@inproceedings{lester-etal-2021-power,
    title = "The Power of Scale for Parameter-Efficient Prompt Tuning",
    author = "Lester, Brian  and others",
    booktitle = "Proceedings of the 2021 Conference on Empirical Methods in Natural Language Processing",
    year = "2021",
}

@inproceedings{oiso24_interspeech,
  title     = {Prompt Tuning for Audio Deepfake Detection: Computationally Efficient Test-time Domain Adaptation with Limited Target Dataset},
  author    = {Hideyuki Oiso and others},
  booktitle = {Interspeech 2024},
  pages     = {2710--2714},
  doi       = {10.21437/Interspeech.2024-81},
  issn      = {2958-1796},
}

@inproceedings{ref5,
  title={Combining evidences from mel cepstral, cochlear filter cepstral and instantaneous frequency features for detection of natural vs. spoofed speech},
  author={Patel, T B and others},
  booktitle={Interspeech},
  year={2015}
}

@inproceedings{jung2022aasist,
  title={Aasist: Audio anti-spoofing using integrated spectro-temporal graph attention networks},
  author={Jung, Jee-weon and others},
  booktitle={ICASSP 2022}
}

@inproceedings{ge2025post,
  title={Post-training for Deepfake Speech Detection},
  author={Ge, Wanying and others},
  booktitle    = {Proceedings of the IEEE ASRU},
  year={2025}
}

@article{ref6,
  title={Constant Q cepstral coefficients: A spoofing countermeasure for automatic speaker verification},
  author={Todisco, M and others},
  journal={Computer Speech \& Language},
  volume={45},
  pages={516--535},
  year={2017}
}

@inproceedings{guo2024audio,
  title={Audio deepfake detection with self-supervised wavlm and multi-fusion attentive classifier},
  author={Guo, Yinlin and others},
  booktitle={ICASSP 2024-2024 IEEE International Conference on Acoustics, Speech and Signal Processing (ICASSP)},
  pages={12702--12706},
  year={2024},
  organization={IEEE}
}

@inproceedings{xuan2025fakemamba,
  title        = {Fake-Mamba: Real-Time Speech Deepfake Detection Using Bidirectional Mamba as Self-Attention's Alternative},
  author       = {Xi Xuan and others},
  booktitle    = {Proceedings of the IEEE ASRU},
  year         = {2025}
}

@inproceedings{baevski2020wav2vec,
  author    = {Alexei Baevski and others},
  title     = {wav2vec 2.0: A Framework for Self-Supervised Learning of Speech Representations},
  booktitle = {Advances in Neural Information Processing Systems (NeurIPS)},
  year      = {2020}
}

@inproceedings{xin2022investigating,
  author    = {Xin W.},
  title     = {Investigating Self-Supervised Front Ends for Speech Spoofing Countermeasures},
  booktitle = {The Speaker and Language Recognition Workshop (Odyssey 2022)},
  year      = {2022}
}

@inproceedings{ref19,
  title     = {{XLS-R: Self-supervised Cross-lingual Speech Representation Learning at Scale}},
  author    = {{Arun Babu and others}},
  year      = {{2022}},
  booktitle = {{Interspeech}},
}

@INPROCEEDINGS{10832361,
  author={Liu, Andy T. and others},
  booktitle={2024 IEEE Spoken Language Technology Workshop (SLT)}, 
  title={Efficient Training of Self-Supervised Speech Foundation Models on a Compute Budget}, 
  year={2024}
}

\end{document}